\documentclass[lettersize,journal]{IEEEtran}
\usepackage{graphicx} 
\usepackage{listings}
\usepackage{xcolor}
\usepackage{xspace}
\usepackage{stfloats}
\usepackage{enumitem}
\usepackage{amsmath}
\usepackage{fancybox}
\usepackage[ruled,lined,linesnumbered,vlined,algo2e]{algorithm2e}
\usepackage{soul}
\usepackage{url}
\usepackage{calc}
\usepackage[many]{tcolorbox}

\usepackage{booktabs}
\usepackage{setspace}
\usepackage{pifont} 
\usepackage{colortbl}
\usepackage{subcaption} 
\usepackage[normalem]{ulem} 
\usepackage{tikz}

\usepackage{algorithm}
\usepackage{algpseudocode}

\definecolor{lightblue}{rgb}{0.85, 0.92, 1.0} 

\usepackage{graphicx} 
\usepackage{listings}
\usepackage{xcolor}
\usepackage{xspace}
\usepackage{stfloats}
\usepackage{enumitem}
\usepackage{amsmath}
\usepackage{fancybox}
\usepackage[ruled,lined,linesnumbered,vlined,algo2e]{algorithm2e}
\usepackage{soul}
\usepackage{url}
\usepackage{calc}
\usepackage[many]{tcolorbox}

\usepackage{booktabs}
\usepackage{setspace}
\usepackage{pifont} 
\usepackage{colortbl}
\usepackage{subcaption} 
\usepackage{tikz}

\definecolor{pink}{rgb}{1,0.078,0.576}  
\definecolor{backgroundcolor}{rgb}{0.95, 0.95, 0.95} 
\definecolor{keywordcolor}{rgb}{0,0,1}  
\definecolor{stringcolor}{rgb}{0.58,0,0.82} 
\definecolor{commentcolor}{rgb}{0.25,0.5,0.37} 
\definecolor{codegreen}{rgb}{0,0.6,0}
\definecolor{codered}{rgb}{0.6,0,0}
\definecolor{codegrey}{rgb}{0.5,0.5,0.5}
\definecolor{backcolour}{rgb}{0.95,0.95,0.92}
\definecolor{codepurple}{rgb}{0.58,0,0.82} 

\lstdefinestyle{javaStyle}{
    language=Java,
    keywordstyle=\color{keywordcolor}\bfseries,
    stringstyle=\color{stringcolor},
    commentstyle=\color{commentcolor}\itshape,
    basicstyle=\ttfamily\footnotesize,
    showstringspaces=false,
    breaklines=true,
    captionpos=b,
    numbers=left,
    numbersep=5pt,
    numberstyle=\tiny\color{codegrey},
}

\lstdefinestyle{mystyle}{
    language=diff,
    commentstyle=\color{codegreen},
    keywordstyle=\color{magenta},
    numberstyle=\tiny\color{codegrey},
    stringstyle=\color{codepurple},
    basicstyle=\ttfamily\footnotesize,
    breakatwhitespace=false,         
    breaklines=true,                 
    captionpos=b,                    
    keepspaces=true,                 
    numbers=left,                    
    numbersep=5pt,                  
    showspaces=false,                
    showstringspaces=false,
    showtabs=false,                  
    tabsize=2
}

\lstdefinelanguage{diff}{
  morecomment=[f][\color{codered}]-,
  morecomment=[f][\color{codegreen}]+,
  morecomment=[f][\color{blue}]{@@}, 
  morecomment=[f][\color{magenta}]{***}, 
  morecomment=[f][\color{violet}]{!}, 
}

\newcommand{\rcf}{\text{VF}\xspace}
\newcommand{\rcfs}{\text{VFs}\xspace}

\newcommand{\tool}{\textsc{VFArchē}\xspace}
\newcommand{\subtool}{Tracker\xspace}

\newcommand{\withpatch}{\textsc{Patch-present}\xspace}
\newcommand{\nopatch}{\textsc{Patch-absent}\xspace}

\newcommand{\ly}[1]{\textcolor{black}{#1}} 
\newcommand{\lyr}[1]{#1}

\begin{document}

\title{\tool: A Dual-Mode Framework for Locating Vulnerable Functions in Open-Source Software}

\author{Lyuye Zhang,
Jian Zhang,
Kaixuan Li,
Chong Wang,
Chengwei Liu,
Jiahui Wu,
Sen Chen,
Yaowen Zheng,\\
Yang Liu
\thanks{The corresponding author is Jian Zhang.\\ 
\indent Lyuye Zhang, Kaixuan Li, Chong Wang, Chengwei Liu, Jiahui Wu and Yang Liu are with College of Computing and Data Science, Nanyang Technological University, Singapore. (zh0004ye@e.ntu.edu.sg, kaixuan.li@ntu.edu.sg, chong.wang@ntu.edu.sg, chengwei.liu@ntu.edu.sg, jiahui004@e.ntu.edu.sg, yangliu@ntu.edu.sg)\\
\indent Jian Zhang is with School of Software, Beihang University, China. (zhangj3353@gmail.com)\\
\indent Sen Chen is with College of Cryptology and Cyber Science, Nankai University, China. (senchen@nankai.edu.cn)\\
\indent Yaowen Zheng is with Institute of Information Engineering, Chinese Acadamy of Sciences, China. (zhengyaowen@iie.ac.cn)
}
}


\markboth{Journal of \LaTeX\ Class Files,~Vol.~14, No.~8, August~2021}%
{Shell \MakeLowercase{\textit{et al.}}: A Sample Article Using IEEEtran.cls for IEEE Journals}

\maketitle

\begin{abstract}
Software Composition Analysis (SCA) has become pivotal in addressing vulnerabilities inherent in software project dependencies.
In particular, reachability analysis is increasingly used in Open-Source Software (OSS) projects to identify reachable vulnerabilities (e.g., CVEs) through call graphs, enabling a focus on exploitable risks.
Performing reachability analysis typically requires the vulnerable function (\rcf) to track the call chains from downstream applications. However, such crucial information is usually unavailable in modern vulnerability databases like NVD.  
While directly extracting \rcf from modified functions in vulnerability patches is intuitive, patches are not always available. Moreover, our preliminary study shows that over 26\% of \rcf do not exist in the modified functions.  
Meanwhile, simply ignoring patches to search vulnerable functions suffers from overwhelming noises and lexical gaps between descriptions and source code. Given that almost half of the vulnerabilities are equipped with patches, a holistic solution that handles both scenarios with and without patches is required.

To meet real-world needs and automatically localize \rcf, we present \tool, a dual-mode approach designed for disclosed vulnerabilities, applicable in scenarios with or without available patch links. \tool is a three-phase framework consisting of (1) CVE Description Expansion: mitigating the lexical gaps; (2) Hybrid Candidate Selection: accurately extending the \rcf and maintaining a short list of candidates without introducing irrelevant noises; (3) Unsupervised
Candidate Ranking: precisely capturing the semantic correlations between CVE descriptions and \rcfs by pairwise ranking. 
The experimental results of \tool on our constructed benchmark dataset demonstrate significant efficacy regarding three metrics, achieving 1.3x and 1.9x Mean Reciprocal Rank over the best baselines for \withpatch and \nopatch modes, respectively. Moreover, \tool has proven its applicability in real-world scenarios by successfully locating \rcf for 43 out of 50 latest vulnerabilities with reasonable efforts and significantly reducing 78-89\% false positives of SCA tools.

\end{abstract}


\section{Introduction}

The security of Open-source Software (OSS) reuse has increasingly drawn attention from the security community and practitioners. Towards the solution, the necessity of Software Composition Analysis (SCA) in enhancing the security of modern OSS projects cannot be overstated. Identifying vulnerabilities merely through lists of dependencies, such as Bill-of-Material approach~\cite{bom,zhao2023software}, is prone to generating plenty of false alarms regarding known vulnerabilities~\cite{imtiaz2021comparative,dann2021identifying,prana2021out,plate2015impact,hu2023empirical,wu2023ossfp}. Often, these vulnerabilities may never pose a real threat as the specific functionalities they compromise are not utilized in the user application, a situation particularly common with indirect dependencies~\cite{wu2023understanding,ponta2020detection,zhang2023mitigating}. 
Consequently, there is a noticeable shift towards function-level and call-graph-based \lyr{reachability analysis} emphasizing only those vulnerabilities that are reachable from the downstream applications~\cite{zhang2023compatible}.

Performing reachability analysis on vulnerabilities requires the identification of the vulnerable function (\rcf)~\cite{wu2024effective,kang2022test, zhang2024vulnerability,wu2023understanding}, which typically refers to the functions that accommodate the exploitable code. If the vulnerable functions are directly or indirectly called by downstream applications, the vulnerability is considered reachable, thus potentially threatening the applications. Identifying \rcf is indispensable for conducting reachability analysis for SCA tools~\cite{snyk,steady,whitesource,owasp,scantist} and academic vulnerability analysis~\cite{wu2023understanding}. 
However, a variety of vulnerability databases that typically disclose Common Vulnerabilities and Exposures (CVEs)~\cite{snykvuldb,githubadvisory,osvinsight,nvd}, 
has not released \rcf for vulnerabilities in their entries. 
Nowadays, \rcf is usually inferred from patches in both academic and industrial worlds~\cite{wu2023understanding,ponta2020detection,zhang2023mitigating} by extracting the modified functions from diffs.
However, patches are available for only about 66\% of vulnerabilities ~\cite{tan2021locating}. 
Furthermore, \rcf is not necessarily the modified functions in the patch but other related functions, a fact revealed by our preliminary study in Section~\ref{sec:prestudy}. 
Therefore, solely relying on modified functions in patches may lead to missing and incorrect \rcf. 
Although we can directly search for \rcf within the entire repository, this approach is limited by a significant number of noisy candidates in the large function corpus, since it overlooks the role of patches as anchors. Therefore, when patches are available, localizing \rcf beyond patches is a critical yet challenging problem.

Meanwhile, localizing \rcf without patches is also crucial considering that around 40\% of vulnerabilities lack patches. The absence of patches poses another challenge, which means that we can only rely on CVE descriptions as the starting point for finding \rcf within the entire repository. Furthermore, we cannot simply use keywords that explicitly point to \rcf. According to our preliminary study, only 8.5\% of CVE descriptions mention \rcf or vulnerable files. 
Simply applying existing techniques for this purpose suffer from various limitations.
Although the well-studied Bug Localization (BL) shares the similar aim of searching source code from reports~\cite{nguyen2011topic,saha2013improving,ye2014learning,wong2014boosting,zhou2012should,lam2015combining,moreno2014use,rahman2018improving,pradel2020scaffle,wu2014crashlocator,koyuncu2019d}, they focus on finding relevant buggy file/functions with abundant information, such as stack trace, structured bug reports, and changesets. Unlike BL, localizing \rcf mostly relies on patches and vulnerability metadata, such as free-text descriptions and library names, which may be prone to information gaps. 
A recent study~\cite{sun2024tracing} on vulnerability metadata supplementation concentrates on tracing vulnerability-related files, offering limited support for \rcf localization.
Traditional approaches, such as Information Retrieval~\cite{tfidf}, relying solely on lexical overlaps, may not suffice, as the description may only have a semantic correlation with the potential \rcf. Moreover, semantic similarity achieved by matching vectors from embedding models, such as CodeBERT~\cite{feng2020codebert}, is susceptible to scalability issues caused by numerous noisy candidates with similar semantics in the source code.

All in all, the ideal tool should accurately extend \rcf based on patches when patches are available, and establish the semantic bridge over the description \rcf when patches are absent. 
In this paper, we propose \tool, a dual-mode framework for automatically identifying \rcf of disclosed vulnerabilities. \tool is designed to handle both scenarios where patch links are either present or absent, which we refer to as \withpatch mode and \nopatch mode, respectively. The framework consists of three main components in successive phases: CVE Description Expansion, Hybrid Candidate Selection, and Unsupervised Candidate Ranking.

\textbf{\textit{Phase 1:}} CVE descriptions frequently omit essential information, such as attack vectors and root causes~\cite{guo2022detecting}, which limits the availability of meaningful semantic clues. To address this issue, we employ the  CVE Description Expansion as the initial step of \tool, which enriches the original CVE descriptions by selectively incorporating conceptual terms from Common Weakness Enumeration (CWE) descriptions~\cite{cwe}.

\textbf{\textit{Phase 2:}} Simply using the description to match all functions in a project for \rcf would suffer from the overwhelming noise of irrelevant functions. Instead, we introduce the Hybrid Candidate Selection to approximate the scope of \rcf by identifying relevant functions. Due to the differing availability of patches in \withpatch and \nopatch modes, we consider them separately yet technically associated:
\begin{itemize}[leftmargin=9pt]
    \item In \withpatch mode, the key is to identify candidate functions that exist outside the patched areas. To this end, based on findings in the preliminary we develop a rule-based \textit{\subtool} that tracks potential candidate functions based on patches and other relevant files using inter-procedural data flow analysis, maintaining a balance between the coverage and the number of candidates.
    \item In \nopatch mode, things become more challenging due to the lack of patches to determine the \rcf scope. To address this, we present a novel idea that transfers knowledge from \withpatch mode by training a \textit{Classifier}. Specifically, we supervisedly fine-tune a CodeBERT model using the results from \subtool on a training dataset of CVEs with patches, enabling us to approximate a similar range of candidates.
\end{itemize}
In this way, the candidate selection process is principally unified across both modes.

\textbf{\textit{Phase 3:}} Due to the lack of a labeled dataset for CVE-\rcf pairs, supervised ranking of candidates is not feasible. Instead, we design an Unsupervised Candidate Ranking approach that leverages LLMs to conduct fine-grained pairwise ranking to avoid the noise from candidates with subtle semantic differences and the inconsistency introduced by similarity scores. Specifically, we first use a Swiss-system tournament~\cite{swisssys} to efficiently identify the most promising candidates with a reduced number of LLM queries. We then apply exhaustive pairwise comparisons among the top-ranked functions to obtain a reliable final ranking. This hybrid method ensures robust candidate selection while maintaining scalability and efficiency. For the comparison of pairs of functions, in-context learning is employed with LLM to assess the relevance of the functions and the CVE.

Since no diverse dataset is publicly available, we manually curated a benchmark of 600 CVEs, ensuring coverage across 85 distinct CWEs. We also collected a training set of 1,000 CVEs with patches for fine-tuning in the candidate selection process. Furthermore, we incorporated four categories of baselines for a fair comparison, including Bug Localization, Information Retrieval, Embedding models, and designated tools. \ly{Moreover, we collected a training set of 1,000 CVEs consisting of $3,376,090$ CVE-function pairs to fine-tune CodeBERT in the \textit{Retrieval} step. }
We conducted extensive experiments on the benchmark, and the results indicate that \tool substantially outperformed baselines in terms of 
\rcf pinpointing. It achieved 1.3x and 1.9x Mean Reciprocal Rank and 1.4x and 1.6x \textit{Manual Efforts@100} over the best baselines in \withpatch and \nopatch modes. The \textit{Recall@10} could reach 96.30\% and 84.00\% in each mode.
Notably, when applying to 50 latest new CVEs with different vulnerable libraries from previously collected CVEs, \tool could successfully localize \rcf for 43 out of 50 vulnerabilities. Based on \rcf of these vulnerabilities, the false positives of SCA tools were substantially reduced, substantiating the generalizability and real-world applicability of \tool.

In summary, this paper makes the following contributions:
\begin{itemize}[leftmargin=9pt]
    \item We conducted a preliminary study that reveals, for the first time, that \rcf is not defined or modified in 26.67\% of patches, indicating that traditional approaches to parsing patch files may generate considerable false negatives.

    \item We propose a novel framework consisting of three steps: CVE Description Expansion, Hybrid Candidate Selection, and Unsupervised Candidate Ranking. This framework captures rich lexical and semantic information to systematically localize \rcf of disclosed vulnerabilities. 
    
    \item We constructed the first benchmark dataset of \rcf for 600 CVEs by manual curation. The evaluation on the benchmark dataset demonstrates the efficacy of \tool. 
    \item We have released our data and the prototype of \tool for further research at our website~\cite{dataset}.

\end{itemize}

\section{Preliminaries and Motivation}
\subsection{Background}
{\textbf{Vulnerable Function (\rcf)}} refers to the function that induces the vulnerability within the vulnerable library, involving the exploitable code logic. That is to trigger the vulnerability, code in \rcf must be executed. Note that in this paper, \rcf does not involve functions along the vulnerable call chains.
To effectively gauge the threat posed by an OSS library, SCA employs code-centric analyses~\cite{ponta2018beyond} like reachability analysis to discern non-reachable vulnerabilities, thus reducing false alarms. In these analyses, if downstream applications have potential call chains to \rcf, i.e. may directly or indirectly call the \rcf, which is considered a critical risky point, the vulnerability is deemed reachable and therefore threatening~\cite{plate2015impact}. If \rcf is not localized, numerous false positives of unreachable vulnerabilities will be reported. Worse, the vulnerability may not be fully fixed, potentially leading to even more severe exploits, as evidenced by multiple real-world cases. For example, the infamous ``Heartbleed'' vulnerability in OpenSSL numbered CVE-2014-0160~\cite{cve-2014-0160}, allowing attackers to read server memory, was not fixed completely. 
CVE-2014-0224~\cite{cve-2014-0224} later addressed the incomplete fix. Apache \texttt{Struts} has a similar situation for CVE-2017-5638~\cite{cve-2017-5638} and CVE-2017-9805~\cite{cve-2017-9805}. Consequently, \rcf is essential in reachability analysis for SCA and vulnerability fixing.

The collection of \rcf has drawn attention, but there is no reliable and efficient way to pinpoint them. As a compromise, Sun et al.~\cite{sun2024tracing} trace the relevant vulnerable files from descriptions, which cannot be applied to reachability analysis. Existing studies~\cite{zhang2023compatible, wu2023understanding,zhang2023mitigating,plate2015impact,ponta2018beyond} conducting reachability analysis assume \rcf are functions modified in the patches. However, as our pre-study will show, \rcf is not necessarily modified in patches.
Although commercial SCA tools, such as Snyk~\cite{snyk}, manually curate \rcf, they require intensive manual effort and are not publicly disclosed. Hence, it is critical to automatically localize the \rcf for disclosed vulnerabilities.

\subsection{Problem Definition}
We aim to automate the process of \rcf localization for disclosed vulnerabilities from CVE entries, such as descriptions, patches, and vulnerability types. To this end, we formulate the problem. Given a disclosed vulnerability $V$ described by a CVE entry associated with a description $desc_V$ and, optionally, a list of patch links $\mathcal{P}_V$. Let $L$ be the vulnerable open-source library, and let $\mathcal{F}_L$ be the set of all functions in $L$. The objective is to find a function $f^*$ such that $f^* \in \mathcal{F}_L$, and $f^*$ is the origin of $V$. A model may produce a mapping as $\phi: V\rightarrow F$ to give all labels of functions $F = (f_1, f_2,...,f_n)$ where $f_j \in (0,1)$ to indicate whether $f_j$ is $f^*$.  This involves analyzing the semantic relationship between $desc_V$ (and $\mathcal{P}_V$ if available) and the source code of functions in $\mathcal{F}_L$.

\subsection{Preliminary Study of \rcf Location}
\label{sec:prestudy}
To understand the locations of \rcf of CVEs, we conducted a preliminary study by pinpointing \rcf based on patches.

\subsubsection{\textbf{Data Preparation and Analysis}}
\label{sec:prestudy data}
\ly{Firstly, we have included the applicable cases from previous work~\cite{kang2022test,wu2024effective}. As Wu et al.~\cite{wu2024effective} did not provide their evaluation dataset, we could only include their cited datasets involving 85 CVEs into our dataset.
However, this dataset is under representative so that we further expanded the dataset by manual curation.
As of Dec 2024, 101,260 CVEs with rich metadata from NVD. Then, we localized and associated patch links with Java files for 1,676 CVEs from both NVD and other sources over the Internet, such as OSV~\cite{OSV} and GitHub Advisory~\cite{githubadvisory}. 
To cover a range of vulnerability types, we concentrated on all 85 unique CWE identifiers~\cite{topcwe}, resulting in 995 CVEs. Beyond the initial 85 CVEs from the related work, we randomly extended them to 600 CVEs associated with patches and vulnerable Java libraries, ensuring at least one from each CWE. These CVEs impacted 222 distinct Java libraries. Next, they were classified into two categories depending on whether the \rcf was modified in the patch.}

\noindent\textbf{Labeling Procedures}
\label{sec:label-data}
For the 515 extended CVEs out of the 600, we manually labeled their \rcf.
\ding{172} Given a CVE, the annotator first understood the vulnerability type from the description and CWE, such as deserialization. 
\ding{173} Based on the vulnerability type and commit messages, the annotator went through each file change to deduce which files are likely to be relevant because there could exist tangled code changes in a patch link including irrelevant files~\cite{wang2019cora}. \ding{174} The annotator then carefully scrutinized the relevant code changes to understand how the diff fixes the vulnerabilities and localize the possible \rcf in deleted and context lines. 
\ding{175} The annotator traced the file path of each \rcf in the current repository. Note that if the \rcf is from other libraries, we labeled the method that directly uses the \rcf in the current repository as \rcf as well due to our target scope of the current vulnerable library. 
\ding{176} In case only the configuration is changed to fix the vulnerability, the annotator localized the methods that refer to the configuration. This step was conducted in an integrated development environment (IDE), such as Intellij~\cite{intellij} as manual tracing is unrealistic in a large code corpus. 
Following the above procedure, the first two authors independently labeled \rcf according to the CVE descriptions, commit messages, and diffs in commits. The third author determined the final results in case of conflicts. 
\vspace{5pt}

\begin{minipage}[t]{0.95\columnwidth}
\centering
\parbox{\textwidth}{%
\small
\colorbox{lightblue}{\parbox{\dimexpr\textwidth-2\fboxsep}{\textbf{\textit{CVE Description:}} Apache ActiveMQ 5.x before 5.13.0 does not restrict the classes that can be serialized in the broker, which allows remote attackers to execute arbitrary code via a crafted serialized Java Message Service ObjectMessage object.}}
}
\begin{lstlisting}[language=diff,style=mystyle,caption=A example from \textit{activemq} where the vulnerable function \texttt{XStream} is replaced in line 2-3,label={lst:exp1},numbers=left]
public XStreamBrokerContext() {
-  XStream stream=new XStream();
+  XStream stream=XStreamSupport.createXStream();
   stream.processAnnotations(SamplePojo.class);
   ...
+public class XStreamSupport {
+  public static XStream createXStream() {
+     XStream stream = new XStream();
+     stream.addPermission(NoTypePermission);
+     stream.allowTypeHierarchy(Collection.class);
      ...
\end{lstlisting}
\end{minipage}

\ly{The labeling results indicated that in the 600 cases, \rcf in 440 cases were labeled at the modified function in the patches and 160 (26.67\%) were not directed modified. There were 74 cases of disagreement during labeling and 56 out of 74 were caused by situations where both True (\rcf in the patch) and False (\rcf not in the patch) occur in single commits. It means that multiple changes in a commit include both \rcf and changes that lead to another \rcf in other files. For such cases, we labeled them as \textbf{NOT} in modified files because they also required further tracking \rcf in other files like cases where \rcf is not in patches.}

After the labeling, we manually extracted the actual \rcf names and classes and summarized the scenarios of \rcf locations. We found that these cases generally belong to 4 scenarios where \rcf is not directly modified in the patches to fix the vulnerabilities. These scenarios include: 
\begin{itemize}[leftmargin=9pt]
    \item \textit{Replaced Methods} (107 cases, 17.83\%): The function invoked with flawed logic is replaced with another function with the same parameters. The replaced method is considered \rcf.

    \item \textit{Replaced Classes} (47 cases, 7.83\%): Similar to the scenario above, the flawed entity is a class for which the constructors and field initiation are considered candidate \rcf.
    \item \textit{Additional Arguments} (80 cases, 13.33\%): The changed part lies in additional arguments of a method invocation. Due to polymorphism~\cite{polymorph}, the actual invoked method is switched to another one. In the method with additional arguments, there usually implements more logic that makes use of the additional arguments to fix the vulnerability.
    \item \textit{Changes of Configurations} (366 cases, 61.00\%): The vulnerability is caused by incomplete configurations before patching. There are two reasons why such cases are considered that \rcf is not modified in the patch. (1) Those methods that directly or indirectly use the configurations are also vulnerable but not tracked by reachability analysis if only the configurations are considered \rcf. These vulnerable methods should also be tracked during the reachability analysis for SCA for better accuracy. (2) The configurations are often defined out of method definitions, such as in \texttt{field} or enclosed by \texttt{static} keyword. Therefore, these methods are also considered as \rcf.

\end{itemize}

\subsubsection{\textbf{Motivating Example}}
Our preliminary study reveals that \rcf is not always modified in patches, and sometimes not even mentioned. We take an example to demonstrate such cases. 
As illustrated in Listing~\ref{lst:exp1}, a diff snippet of a patch link~\cite{motivatingexmaple} for CVE-2015-5254~\cite{cve20155254} and the description is presented. 
\texttt{ActiveMQ} used to use \texttt{XStream} to process annotations, but the default \texttt{XStream} is vulnerable without protective measures, thus it was replaced with a customized \texttt{XStreamSupport} to tweak the issue. According to our definition, \texttt{XStreamBrokerContext} is considered \rcf.  Unfortunately, there exists another reference cite \texttt{XmlMessageRender}~\cite{motivatingexampleref} using \texttt{XStream} in a vulnerable way. We explored the 23 downstream artifacts on Maven Central~\cite{mvnrepo} and found that $3$ of them had indirect invocation to the vulnerable method \texttt{XmlMessageRender}. Therefore, locating \texttt{XStream} instead of \texttt{XStreamBrokerContext} as \rcf could avoid false negatives of the reachable vulnerabilities.

\noindent\textbf{Whether CVE descriptions directly mention \rcf?} Naturally, one might wonder whether we can simply search for function names in a project when files or functions are mentioned in the CVE description. To explore this, we manually examined the above-mentioned 600 CVEs by searching files and functions in their descriptions. We initially found that 135 of these CVEs mentioned files or functions. However, upon manual verification, we discovered that the function names in 84 of these CVEs were misleading as they did not belong to the vulnerable libraries. These names were mentioned either because they were affected or could contribute to the exploits. Of the remaining 51 CVEs, 27 had vulnerable files mentioned in the description, and 24 had \rcf explicitly identified. This means that only 8.5\% of CVEs had relevant files or functions in their descriptions, and only 4\% specifically identified \rcf. Worse, 14\% of CVEs contained misleading function names in their descriptions. This highlights the need for dedicated approaches to address this challenging problem.

\section{Methodology}
\begin{figure*}[!t]
\centering
  \includegraphics[width=0.9\linewidth]{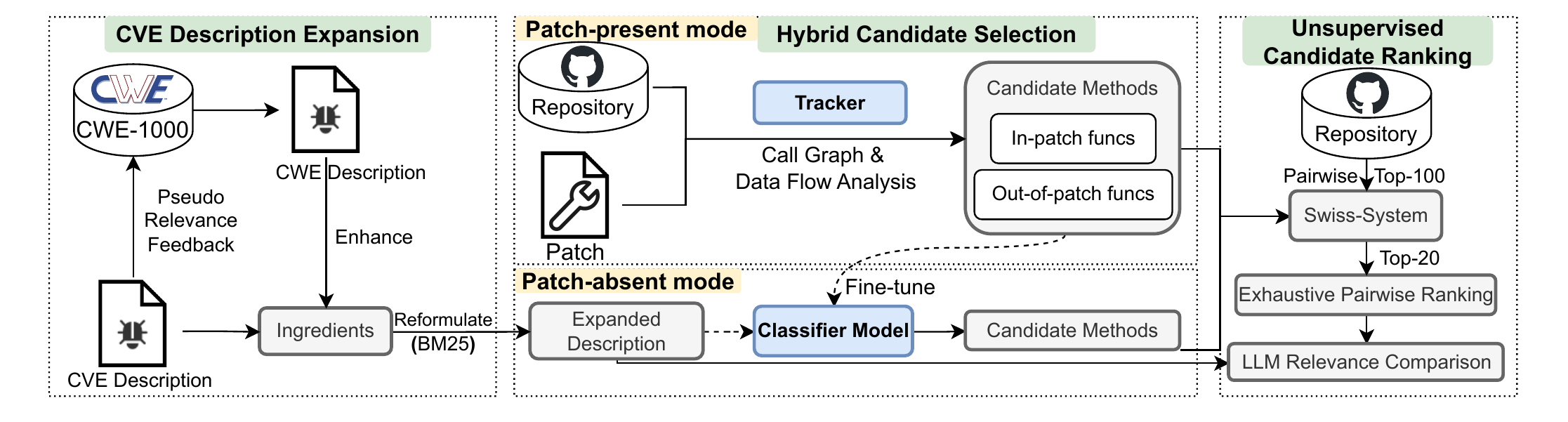}
  \caption{Overview of \tool}
  \label{fig:overview}
\end{figure*}

We propose a three-phase framework to localize \rcf, including \textit{CVE Description Expansion}, \textit{Hybrid Candidate Selection}, and \textit{Unsupervised Candidate Ranking}. The vulnerability descriptions are first expanded with CWE descriptions~\cite{cwe}, as illustrated in Figure~\ref{fig:overview}. 
Next, \tool handles two scenarios where patches are either available or unavailable, namely, \nopatch and \withpatch modes.
When the patch is available, a sub-tool, \subtool, selects all candidate functions from methods or classes involved in patches. 
When the patch is unavailable, we fine-tune a CodeBERT model~\cite{feng2020codebert}, adapting it to leverage knowledge transferred from \withpatch mode. 
Finally, the selected candidates are ranked based on semantic understanding using a chat-based language model.

\subsection{CVE Description Expansion}
Considering that the quality of CVE description varies substantially~\cite{guo2021key}, to compensate for the inadequate description, \tool first expands the description with state-of-the-art QE techniques~\cite{azad2019query}. Note that this step is applied regardless of the presence of patches. As in CWE documentation~\cite {cweguide}, CWE description illuminates the root cause of vulnerabilities with a majority of CVEs associated. 
Thus the ingredients used for expansion are from the CWE descriptions if available. 
In cases where no CWE is associated with the CVE, we use the Pseudo Relevance Feedback (PRF) to retrieve the most relevant CWE description from all descriptions provided in CWE-1000~\cite{cwe}. Specifically,  we leverage the BM25 algorithm frequently used in PRF~\cite{wang2023query2doc} to refine CVE descriptions by incorporating the most pertinent CWE descriptions. PRF iteratively enhances the original query by assuming the top-ranked documents from an initial retrieval are relevant. Specifically, we apply the following PRF formula
$Q' = Q + \alpha \sum_{d \in D} \text{BM25}(d, Q) \cdot \text{Vector}(d)$
where $Q'$ is the expanded query. $d$ is the description and $\alpha$ is the co-efficient.

Upon deriving the ingredient, the second step is reformulation (a.k.a term re-weighting)~\cite{azad2019query}, which integrates  Latent Semantic Analysis (LSA)~\cite{lsa} with Term Frequency-Inverse Document Frequency (TF-IDF)~\cite{tfidf} vectorization. This process begins with English word sanitization, such as stop word removal and lemmatization of the corpus with NLTK~\cite{nltk}.
Subsequently, the corpus is transformed into a TF-IDF matrix, emphasizing the importance of terms.

The application of LSA, through Singular Value Decomposition (SVD) on the TF-IDF matrix, facilitates dimensionality reduction, capturing latent semantic associations within the corpus. This is formally represented as $A = U\Sigma V^T$, where $A$ is the TF-IDF matrix, $U$ and $V$ are orthogonal matrices, and $\Sigma$ is a diagonal matrix of singular values. The transformation into a reduced-dimensional semantic space allows for the semantic projection of both the query and candidate expansion terms.

Semantic similarity between the projected query vector and each candidate term vector is assessed using cosine similarity, which serves as the weights for the expanded terms. Finally, the candidate expansion terms are deduplicated to ensure the uniqueness of terms in the expanded query. Apart from the expanded part, we repeated the original query for 5 times to emphasize it according to common practices~\cite{wang2023query2doc}.

\subsection{Hybrid Candidate Selection}
Based on the expanded description, a subset of candidate functions (Java methods) is selected from the repository, setting the stage for further refinement in the subsequent step. Note that the \texttt{constructor} is deemed as a special method named by the \texttt{class} name. 
As the presence of patch links serves as a critical anchor, our retrieval adopts a hybrid method that incorporates two distinct modes depending on the availability of patches: \withpatch mode and \nopatch mode.

\subsubsection{\textbf{\withpatch Pattern-based \subtool}} 
According to the pre-study in Section~\ref{sec:prestudy}, although \rcf is not necessarily modified in the patch, it is always directly or indirectly related to functions modified in the patch. Thus they can be tracked either by invocation or data flow. Based on the patch and the corresponding source code repository associated with the CVE, we developed \subtool to extend potential candidate methods based on call graph and data flow analysis. 

Ideally, \subtool should verify the existence of \rcf within the patch before widening the candidate scope. However, it is not feasible to perfectly identify such information, since it demands a human-level understanding of the CVE description, commit messages, and the code context, alongside a security background.
Thus, a compromise was made by sacrificing precision to ensure \rcf is not neglected by \subtool. \subtool conducts pattern matching to iteratively identify relevant methods from modified methods or classes in the patch. Although this process might introduce some noise, the subsequent \textit{Ranking} phase can tolerate it by applying heavy but more accurate semantic similarity measurement. Specifically, \subtool leverages JavaParser~\cite{javaparser} to perform source-code-level program analysis. \ly{Although the patterns are designed for Java, the strategy can be migrated to other programing languages by adjusting the patterns.} For each scenario in the Pre-study~\ref{sec:prestudy}, \subtool customizes a pattern to identify: 
\begin{itemize}[leftmargin=9pt]
    \item \textit{Replaced Methods}: In a hunk of patch, if both the deleted line and the added line involve method invocations and the parameters of them are within the same Def-Use chain, then it is deemed as \textit{Replaced Methods}. The deleted method name is used to obtain the source code in this repository. The alias analysis of the parameters is committed by tracking the data flow to ensure the detection is robust against refactoring. 
    \item \textit{Replaced Classes}: This procedure is similar as the above. The difference is that both class initialization and invocation of Static methods are considered. For example, in Listing~\ref{lst:exp1}, the deleted line is class initialization, and the added line is Static method invocation. 
    \item \textit{Additional Arguments}: This scenario includes both the class and the method. If an invocation or an initialization appears in both deleted and added lines in a hunk but with different parameters, it is recorded. The method or class names are saved to track the source code. Note that the parameters aliases are also normalized against refactoring.
    \item \textit{Changes of Configurations}: Due to Java-specific features, globally accessible variables are usually defined in class fields. Hence, \subtool tracks the field changes. Note that, apart from field definitions, the code enclosed by keyword \texttt{Static} in a Java class is also tracked because Java allows the static way to initialize the field with \texttt{Static}. After deriving the target field names, \subtool iteratively searches for the references of these field names in other files of the repository with data flow analysis. Names of the methods that refer to these field names are collected. 
    \item \textit{No pattern matches}: In case no pattern is matched, \subtool returns the modified methods/classes in the patch, meaning \subtool determines that \rcf is modified in the patch.
\end{itemize}

Based on the above patterns, the source code body of the candidate method/class is derived for \textit{Ranking} in the next step. 

\begin{algorithm2e}[t!]
    \footnotesize
    \SetAlgoNlRelativeSize{0}
    \SetNlSty{textbf}{(}{)}
    \SetCommentSty{mycommfont}
    
    \caption{\subtool for Identifying \rcf Candidates}
    \label{alg:1}
    \DontPrintSemicolon
    
    \KwIn{Patch diff $Diff$, Target Repository $Repo$}
    \KwOut{List of candidate methods $Method_{cand}$}
    $Hunk \gets Diff.split()$\;
    $File \gets Repo.walk().filterTesting()$\;
    \ForEach{$h \in Hunk$}{
        $added, deleted, context \gets h$\;
        $Method_{add} \gets extractSig(added)$\;
        $Method_{del} \gets extractSig(deleted)$\;
        \ForEach{$m_{del} \in Method_{del}$}{
            \If{$m_{del} \in Method_{add}$}{
                $args_{del} \gets normalizeAlias(m_{del}.args)$\;
                $args_{add} \gets normalizeAlias(m_{add}.args)$\;
                \If{$args_{del}$ = $args_{add}$ and $m_{del}\in File$}{
                    $Method_{cand}.add(m_{del})$\;
                }
            }
            \If{$m_{del}.name = m_{add}.name$}{
                $args_{del} \gets normalizeAlias(m_{del}.args)$\;
                $args_{add} \gets normalizeAlias(m_{add}.args)$\;
                \If{$args_{del} \neq args_{add}$ and $m_{del}\in File$}{
                    $Method_{cand}.add(m_{del})$\;
                }
            }
        }
        \If{$context \text{is \texttt{field}} \text{\textit{ or }} context \text{is \texttt{static}}$}{
            $Field \gets added.variables \cup deleted.variables$\;
            \ForEach{$field \in Field$}{
                \ForEach{$method \in File.extractMethod()$}{
                    \If{$method.reference(field)$}{
                        $Method_{cand}.add(method)$\;
                    }
                }
            }
        }

    }

    \KwRet{$Method_{cand}$}
\end{algorithm2e}

\subsubsection{\textbf{\nopatch Learning-based Classifier}}
When the patch link is unavailable, accounting for 40-55\% of CVEs~\cite{tan2021locating}, there is no anchor for \subtool to perform. Because directly applying \textit{Ranking} to functions of the entire project involves a plethora of irrelevant functions, the retrieval of candidate functions is also needed for \nopatch mode. 
Thus, we transfer the ability of \subtool to the \nopatch mode by fine-tuning a pre-trained model, CodeBERT~\cite{feng2020codebert}, with pairs of candidate methods from \subtool and CVE descriptions to classify methods solely based on descriptions without patches.

To effectively identify VOF candidates in \nopatch mode, we utilize the cross-encoder architecture of CodeBERT, fine-tuned for our specific task. This approach stands out for its ability to process the interaction between CVE descriptions and Java method tokens, providing a bimodal understanding crucial for our classification task. The cross-encoder architecture is optimally suited for our task since it also aligns with the format of comprehensive pretraining on extensive code and natural language datasets. Therefore, this training enables the model to fully leverage the domain knowledge of CodeBERT for determining whether a function is within the CVE scope. 

For preparing the fine-tuning dataset,  we initially executed \subtool on repositories containing vulnerable libraries. These repositories were associated with a set of 1,000 randomly selected Java CVEs with corresponding patches. Importantly, these CVEs were distinct from those considered in our preliminary study. This initial step yielded a total of $33,944$ Java methods. Afterward, to accurately simulate scenarios where patches are present, we calculated the average ratio of candidate methods to all non-testing methods based on the pre-study dataset. The obtained ratio was $1:107.452$, reflecting the distribution of candidate methods within the scope of our study. For practicality in dataset construction, we approximated the ratio to $1:100$ for assembling the target negative dataset. This decision led to the selection of $3,342,146$ methods to form the negative dataset. Consequently, the fine-tuning dataset was comprised of a total of $3,376,090$ entries.

Formally, our dataset $\mathcal{D}$ consists of pairs $(d, m)$, where $d$ is a CVE description, and $m$ represents a method, with each pair labeled $y \in \{0,1\}$ to denote relevance. The input to CodeBERT expressed as $X = d \oplus m$, concatenates $d$ and $m$. CodeBERT processes the concatenated input $X$ through its self-attention mechanism, allowing each token in the input sequence to interact with every other token. This interaction is key to understanding the contextual relationships within the input, allowing the model to capture the semantics of both the CVE description $d$ and the Java method $m$. The self-attention mechanism produces a set of output vectors, one for each input token, including a special output vector for the [CLS] token at the beginning of the sequence. This vector, denoted as $\mathbf{v}_{\text{[CLS]}}=f(X)$, encapsulates the global understanding of the entire input sequence, representing $d$ and $m$. 

The vector $\mathbf{v}_{\text{[CLS]}}$ is then utilized as the input to a linear layer, which is represented as follows:
\[
p = \sigma(\mathbf{W} \cdot \mathbf{v}_{\text{[CLS]}} + b)
\]
Here, $\sigma$ denotes the sigmoid activation function, $\mathbf{W}$ represents the weight matrix of the linear layer, and $b$ is the bias term. The resulting score $p$ indicates the probability of relevance of the Java method $m$ to the CVE description $d$, leveraging the comprehensive context captured by $\mathbf{v}_{\text{[CLS]}}$.

The fine-tuning objective is to minimize the binary cross-entropy loss $\mathcal{L}$ across all pairs in the dataset $\mathcal{D}$, with the loss for $n$ pairs defined as:
$$
\mathcal{L}(\theta) = -\frac{1}{n}\sum_{i=1}^{n} \left[ y_i \log(p_i) + (1 - y_i) \log(1 - p_i) \right]
$$
This optimization refines the model parameters $\theta$, thereby improving the model's capability to accurately classify Java methods in relation to CVE descriptions, enhancing the process of VOF candidate identification without patches.

In the candidate selection phase, given a CVE alongside the repository, we construct $X'$ for each pair consisting of the CVE description and a non-testing method from the repository. For each pair, the fine-tuned model $f$ predicts the likelihood $p_i$, resulting in a series of scores ($p_1, p_2, \ldots, p_n$) for $n$ pairs. These scores are then ranked in descending order to prioritize the methods most likely to be within the scope of \rcf for the CVE, allowing for the selection of the top 100 methods. This process yields a subset of methods as candidates that effectively simulate the \withpatch scenario. As such, we establish a unified retrieval phase to seamlessly integrate with the ranking process in the next section.

\subsection{Unsupervised Candidate Ranking}

Traditional methods that rank candidate functions based solely on similarity scores (e.g., cosine similarity between embeddings) often fail to capture the nuanced and context-dependent relationships between CVE descriptions and candidate functions. These methods rely on the assumption of a globally consistent score distribution, which rarely holds in the presence of complex code semantics or diverse vulnerability patterns. Moreover, the lack of a large labeled dataset of \rcf pairs makes supervised fine-tuning infeasible. To address these challenges, we adopt an unsupervised pairwise ranking approach.

To address these limitations, we employ a \textbf{pairwise comparison} paradigm based on in-context learning (ICL), where LLMs are provided with relevant context and examples directly in the prompt. The ranking first leverages the Swiss-system Tournament~\cite{swisssys} to produce a top-20 from top-100, then runs the exhaustive ranking for better reliability. For each pair of comparisons, ICL enables the LLM to adaptively reason about the relative relevance of two candidate functions with respect to the given CVE, without the need for resource-intensive fine-tuning.
This approach allows for a more robust and context-aware ranking, as the LLM can integrate both the CVE context and the specific characteristics of the candidate functions, going beyond fixed-vector similarities and incorporating richer semantic understanding. 

\subsubsection{Candidate Pool Reduction}
Let $\mathcal{C} = \{f_1, f_2, \dots, f_N\}$ denote the set of $N$ candidate functions obtained after initial filtering ($N \leq 100$). Due to the prohibitively large number of pairwise comparisons required for all $N = 100$ candidates, the total number of comparisons is given by the formula $\frac{N(N-1)}{2}$, resulting in $4,950$ distinct comparisons. This leads to a significant computational and financial cost when querying LLMs. To alleviate this inefficiency while minimizing the loss in ranking quality, we adopt the Swiss-system tournament~\cite{swisssys}, which strategically schedules matches to focus on the most informative pairings. By eliminating redundant comparisons between pairs with evident differences in ranking, the Swiss system reduces the total number of comparisons to approximately $\frac{N}{2} \times R$, where $R$ is the number of rounds. For $N = 100$ and $R = 8$ (typical for $N=100$), this approach requires only about $400$ LLM queries, representing a substantial reduction in comparison overhead.
Empirically, we set $R = 8$ for $N \approx 100$, which balances ranking reliability and efficiency. Each comparison awards $1$ point to the winner, $0$ to the loser, and $0.5$ to each in the case of a tie.

\subsubsection{Pair-wise Comparison for Ranking}
While the Swiss-system tournament efficiently narrows down the candidate pool, it does not guarantee that the most relevant function will be ranked at the very top, but rather ensures it is likely to be among the highest ranks. To further enhance ranking reliability for the most promising candidates, we conduct exhaustive pairwise LLM comparisons within the set of top-ranked functions, denoted as $\mathcal{C}_{\text{top}}$. This hybrid ranking strategy, which combines the efficiency of the Swiss-system tournament with the accuracy of exhaustive pairwise ranking in the final stage, achieves a practical balance between computational efficiency and ranking reliability.
Let $S_i$ denote the Swiss-system score for candidate $f_i$ at the end of the tournament:
\[
S_i = \sum_{j=1}^{R} s_{i,j}
\]
where $s_{i,j} \in \{0, 0.5, 1\}$ is the score from $f_i$ in round $j$.

After computing all $S_i$, we select the top $K$ candidates $\mathcal{C}_{\text{top}} = \{f_{(1)}, \dots, f_{(K)}\}$ with the highest $S_i$, where $K=20$ in our setting.

For each unordered pair $(f_i, f_j) \in \mathcal{C}_{\text{top}}$, we query the LLM to determine which function is more likely to be \rcf. The result for each candidate $f_i$ is a pairwise win count:
\[
W_i = \sum_{j \neq i} I(f_i \succ f_j)
\]
where $I(f_i \succ f_j) = 1$ if $f_i$ is preferred over $f_j$, $0.5$ in the event of a tie, and $0$ otherwise.
The final ranking is determined by sorting $\mathcal{C}_{\text{top}}$ in descending order of their pairwise win counts $W_i$. Each candidate $f_i$'s final score used for ranking is $W_i$.

Each pairwise comparison is performed using an LLM under the ICL paradigm, where relevant context and example prompts are provided to guide the model’s reasoning. The LLM's output is then mapped to the corresponding comparison score as described previously. To ensure consistency and efficiency, we cache the comparison results by storing a hash of the CVE description and candidate pair, thereby avoiding redundant queries and ensuring reproducibility of the ICL-based judgments.

This hybrid approach is designed to maximize ranking reliability and MRR for the most relevant candidates while keeping the number of LLM queries manageable. By first applying the Swiss-system tournament to narrow down the candidate pool, \tool focuses computational resources on the most promising subset. Exhaustive pairwise LLM comparisons within this subset then provide a precise final ranking. The use of in-context learning in each LLM query further enhances ranking accuracy by allowing the model to leverage relevant contextual information and demonstration examples.

\section{Experimental Design}
In the evaluation, we first verify \tool's effectiveness in a benchmark dataset, along with a comparison with other baselines. Then, an ablation study is conducted to embody the contribution of individual parts of \tool. Lastly, an application study is performed to understand the generalizability of \tool for the latest vulnerabilities in different libraries from benchmark and training datasets. 
To this end, three research questions (RQs) are raised:

\begin{itemize}[leftmargin=9pt]
    \item \textbf{RQ1 (Effectiveness)}: How effective is \tool in \textbf{\withpatch} and \textbf{\nopatch} modes?
    \item \textbf{RQ2 (Ablation Study)}: What is the contribution of each individual component within \tool?
    \item \textbf{RQ3 (Application Study)}: How does \tool perform regarding localizing \rcf and enhancing existing SCA tools for the latest vulnerabilities in real-world scenarios over baselines?
\end{itemize}

\subsection{Data Preparation}
\label{sec:eval-data}
Due to the absence of a publicly available \rcf dataset, based on the dataset from a related study~\cite{wu2024effective}, we manually curated and open-sourced a benchmark dataset for future research. We incorporated the 600 Java vulnerabilities from the Pre-study in Section~\ref{sec:label-data} into the dataset. The elaboration of labeling is introduced in Section~\ref{sec:label-data}.

\subsection{Implementation}
For the supervised training of CodeBERT using 1,000 CVEs with $3,376,090$ records, we split the dataset into a training set, a validation set, and a testing set in a proportion of $8:1:1$. The maximum token length is set as 512 by default, which is mostly enough to encapsulate the entire method. The English word sanitization involves the NLTK~\cite{nltk} and Ronin splitter~\cite{Hucka2018} as well as the Roberta tokenizer of CodeBERT. 
During fine-tuning,
the learning rate is configured as 0.00005 with 3 epochs based on the recommendations in the original BERT paper~\cite{feng2020codebert} and has generally been found to work well for similar fine-tuning tasks. A total of 500 steps are designated as warmups, allowing for a gradual ramp-up of the learning rate to its maximum value, mitigating the risk of early training instability. Additionally, to prevent overfitting, a weight decay regularization of 0.01 is applied, encouraging the model to learn simpler, more generalizable patterns. The Hybrid Ranking uses GPT-4.1~\cite{gpt41} as the base model for pair comparison. All the experiments ran on a server with 80 CPU cores (Intel(R) Xeon(R) CPU E5-2698 v4 @ 2.20GHz), 512GB RAM, and 8 NVIDIA Tesla V100-SXM2 (32 GB memory each), operating on GNU/Linux Ubuntu 20.04.6.

\subsection{Baselines}
\label{sec:baseline}

\ly{The baselines cover two modes of patch presence. For \withpatch mode, since no existing tool considers patches to localize \rcf, we developed a straightforward baseline using TF-IDF to rank only modified functions in patches as $\text{TF-IDF}_{p}$. For \nopatch mode, to conduct a comprehensive comparison, besides designated tools, we introduced and included tools that could perform similar tasks by searching source code files based on natural language descriptions, such as Bug Localization tools, Information Retrieval (IR) tools, and Embedding Models. In total, four categories of baselines were involved:} 
\ly{
\begin{itemize}[leftmargin=9pt]
    \item For \textbf{Bug Localization}, many approaches~\cite{ye2014learning,wong2014boosting,zhou2012should,lam2015combining,moreno2014use,pradel2020scaffle, wu2014crashlocator,koyuncu2019d, lam2017bug} have been proposed, yet they rely on stack trace or changesets which are not available in our scenario. Hence, we only focused on those designed for source code. Although some, including BugScout~\cite{nguyen2011topic}, BLUiR~\cite{saha2013improving}, can work on bug reports, they were published around 10 years ago and artifacts are unavailable. 
    We included the latest Bug Localization tool SemanticFlowGraph~\cite{fse-SFG} (SFG) and Blizzard~\cite{rahman2018improving} as baselines. SFG as a pre-trained model is designed for both source code and changesets, thus we only compared its source code mode.
    \item  For \textbf{Information Retrieval}, BM25 as the advanced technique over TF-IDF was included. 
    \item For \textbf{Embedding Models}, 
    Openai embedding model~\cite{openaiembedding}, a SOTA embedding model, is employed in the comparison where \textit{text-embedding-3-small} was adopted. In \withpatch mode, it is named $openai_{p}$, and $openai_{a}$ is used in \nopatch mode.
    CodeBERT~\cite{feng2020codebert}, an encoder-only model, has been widely used as the embedding model for code and understanding, such as tracking bugs from changesets by Ciborowska et al.~\cite{ciborowska2022fast}, thus CodeBERT-base~\cite{codebertbase} was included as a baseline. UniXcoder~\cite{guo2022unixcoderunifiedcrossmodalpretraining} and Word2Vec~\cite{mikolov2013efficient} are widely adopted models to track down buggy code by measuring semantic relationships between texts and code~\cite{zou2021bleser}. Hence, we included pre-trained UniXcoder-base~\cite{unixcoder} and Word2Vec-base~\cite{w2vbase} as baselines by calculating the vector representations of descriptions and source code that are sanitized with the same approach as \tool. Then, the cosine similarities for each method are ranked in descendingly order. 
    \item For \textbf{Designated Tools} designed specifically for vulnerability data supplementation. So far, there are two recent SOTA tools in 2024, a vulnerable file tracking tool~\cite{sun2024tracing} (abbreviated as VulF in this paper) 
    and VFFinder~\cite{wu2024effective}, a vulnerable function tracking tool. 
    Although we have attempted to contact VFFinder's authors and reproduce VFFinder, its artifact is still not runnable by the time of our paper submission with multiple absent critical components. 
    Since it leverages natural language sequences, such as CVE descriptions, to localize the vulnerability-related files with training scripts provided, we trained its base model, transformer~\cite{tensor2tensor}, based on our dataset and performed the evaluation. 
\end{itemize}
}

\subsection{Evaluation Metrics}
We adopted three commonly used metrics to comprehensively evaluate \tool and baselines: \textit{Recall@K}, Mean Reciprocal Rank (MRR)~\cite{gu2018deep}, and \textit{Manual Efforts@K}. 

\textbf{Recall@K} is used to measure the ability of ranking systems to retrieve all relevant items within the top-K ranked predictions. Given top-K predictions, it refers to the proportion of the number of \rcfs over the total number of \rcfs. It is defined as:
$Recall@K = \frac{|S_{q,K} \cap R_q|}{|R_q|}$
\noindent where \(|S_{q,K} \cap R_q|\) is the number of relevant items in the top \(K\) predictions. \(|R_q|\) is the total number of relevant items for \(q\).

\textbf{Mean Reciprocal Rank (MRR)} has been extensively adopted to measure the capability of ranking systems to rank the first relevant element.
MRR is defined as: 
$MRR = \frac{1}{|Q|} \sum_{i=1}^{|Q|} \frac{1}{rank_i}$
where $|Q|$ is the total number of queries and $rank_i$ is the rank position of the first \rcf for the $i^{th}$ query.

\textbf{Manual Efforts@K (ME@K)} has been increasingly adopted in recent studies~\cite{tan2021locating,wang2022vcmatch} as a great indicator of the manual inspection efforts to verify a ranked list. For top-K elements for each query, the effort is calculated as the rank of the first relevant element up to K (maximal efforts). It is defined as:
$ME@K = \frac{1}{|Q|} \sum_{q=1}^{|Q|} \min\left(\frac{p_q}{K}, 1\right)$
\noindent where \(|Q|\) is the total number of queries. \(p_q\) is the rank position of the first correct answer within the top \(K\) predictions for the \(q^{th}\) query. 
\(K\) is the predefined cutoff rank, indicating the limit beyond which predictions are not considered.

\section{Results and Analysis}
\subsection{RQ1: Effectiveness}
\label{sec:rq1}
Dual modes of \tool are assessed regarding effectiveness.
\subsubsection{\textbf{\withpatch Mode}}

In Table~\ref{tab:benchmark}, \tool achieved an MRR of 0.78, which indicates that \tool mostly ranks the first relevant method at a high position. For $Recall@K$, although \tool only achieved $67\%$ for $Recall@1$, it swiftly climbed to 96.3\% for $Recall@10$. This is because some cases include multiple \rcfs, especially for cases where \rcf is not modified in patches, e.g., \textit{Changes of Configurations}. Consequently, $Recall@1\text{-}3$ may not capture all relevant methods due to the limited window size $k$. Nevertheless, with the expansion of \(k\), the majority of \rcf find their place within the top-10 ranked list. Regarding \(ME@K\), it is noteworthy that the average \(ME\) remains relatively stable across \(ME@1\text{-}100\), with \(ME\) only reaching \(4.88\) even at \(ME@100\). 
This stability suggests that \subtool only tracks down the relevant methods from the patch without incurring much noise.

As a point of comparison, $\text{TF-IDF}_{p}$ achieved inferior results because it does not track potential \rcf from patches, relying solely on modified functions, which leads to false negatives. However, the number of false positives is relatively low for $\text{TF-IDF}_{p}$, as it generates a smaller rank list from within the patches compared to \tool. 
Its performance is capped due to the inability to identify \rcfs outside of the patches, as our preliminary study has shown.

The OpenAI embedding model $Openai_{p}$ achieves promising results, as it is a more advanced large embedding model compared to traditional models such as CodeBert. However, it still does not outperform \tool, since it does not leverage pairwise ranking or domain-specific cues used by \tool.

Compared with \tool in \nopatch mode, \tool with patches could achieve generally better results in all three metrics, thanks to \subtool. This is because the patches can serve as the anchor to effectively eliminate irrelevant methods. 
At \textit{Candidate selection} step, on average, \tool returned $17.61$ candidate methods per CVE in \withpatch mode. In contrast, in \nopatch mode, the number is $99.21$ as the classifier model always returns the top 100 methods.

\subsubsection{\textbf{\nopatch Mode}}
The evaluation was conducted on the same dataset without providing patches to \tool and other tools. The efficacy of \tool in handling \nopatch CVEs is highlighted in the right section of Table~\ref{tab:benchmark}. With an MRR of 0.58, \tool significantly surpasses the performance of baselines. This superiority is attributed to the absence of a \textit{Candidate selection} step in other tools, which forces them to process an excessive number of noisy methods, thereby compromising their effectiveness. This issue similarly affects other metrics, with \tool outperforming its counterparts. Particularly notable is \tool's performance in $Recall@K$ with a rapid improvement. In contrast, the $Recall@K$ for other tools increases at a much slower pace, remaining suboptimal (around $30\%$) at $Recall@10$.

Among the baselines, the OpenAI embedding model achieved better results than the other baseline methods. However, in the \nopatch mode, there are typically around 100 candidates, which introduces more noise compared to the \withpatch mode. As a result, it becomes more challenging to rank the correct candidates at the top, leading to the inferior performance of the embedding model under the pairwise ranking approach. This is because the model is more susceptible to subtle noise within the candidate set. Even the SOTA text embedding model, the OpenAI, was unable to achieve results comparable to those of \tool, which incorporates a pairwise ranking algorithm.

BM25 achieved better MRR yet slightly worse $Recall@K$ and $ME@K$. As the representative of IR techniques, BM25 depends on term statistics, lacking semantic insight. This limits their effectiveness when terms in the code and description do not match. For example, if a description mentions \textit{deserialization} but the code never uses this term, IR might miss relevant methods. Hence, the diversity of our benchmark makes IR alone struggle to rank \rcf.

For pre-trained models including Word2Vec, UniXcoder and CodeBERT, Table~\ref{tab:benchmark} reveals superior performance of CodeBERT over Word2Vec in all metrics. This discrepancy in performance can be largely attributed to CodeBERT's pretraining on extensive source code and accompanying documentation, a step not undertaken by Word2Vec. Additionally, CodeBERT's self-attention mechanism more effectively aligns the semantics between source code and documentation. 
\ly{UniXcoder as the more advanced model than CodeBERT indeed provided slightly better performance in almost all metrics.
While MRRs of both UniXcoder and CodeBERT were lower compared to IR techniques, $ME@K$ of them grows gently with $K$, meaning the relevant methods were mostly ranked relatively higher in the top 100 than IR-based BM25. They provided a better potential for ranking all relevant candidates in high order.}

\ly{SFG is the latest Bug Localization tool that specifically captures the semantics from code by leveraging the code structure information. It achieved a relatively great performance regarding most metrics. This pre-trained model was built on top of CodeBERT by enriching the structural semantics, resulting in much better effectiveness than CodeBERT-base.
}
Unexpectedly, Blizzard, which typically excels in Bug Localization tasks, performed lower than all other tools in the \rcf localization scenario. In particular, its MRR was only 0.04, with similar results observed across other metrics. This may primarily be blamed on the absence of stack traces, which are typically available and useful in Bug Localization.
Furthermore, our inspection of Blizzard's queries expanded by its own QE revealed disorganized text filled with non-English words. Its QE results in queries filled with noisy terms that do not align with \rcfs, leading to suboptimal performance in \rcf localization.

\ly{VulF~\cite{sun2024tracing}, as a dedicated tool, demonstrated the highest performance among the baseline methods, though it was still outperformed by \tool. Despite VulF’s original design at the file-based granularity rather than the function level, its performance could potentially improve through fine-tuning with a function-based dataset, leveraging its inherent sequence-based transformer architecture. Nevertheless, without a candidate screening phase, such as the \textit{Retrieval} step in \tool, ranking all functions introduces excessive noise, resulting in sub-optimal effectiveness.}

\begin{tcolorbox}[boxrule=0.5pt,arc=1pt,boxsep=-1mm,breakable]
\textbf{Answer RQ1}: 
In \withpatch mode, \tool achieves an MRR of 0.78 and maintains high $Recall@K$, with $ME@K$ remaining low as $K$ increases. In \nopatch mode, \tool secures a 0.58 MRR, outperforming IR, embedding models, and Bug Localization tools. It reaches a $Recall@10$ of 96\% and $ME@10$ of 2.03, efficiently ranking the \rcf within the top 10.
\end{tcolorbox}

\begin{table*}[]
\small
\setlength{\tabcolsep}{4pt}
\caption{Effectiveness Comparison on Benchmark Dataset among Baselines in Dual-mode}
\label{tab:benchmark}
\begin{tabular}{l|rrr|rrrrrrrrr}
\hline
\rowcolor[HTML]{FFFFFF} 
Mode      & \multicolumn{3}{c|}{\cellcolor[HTML]{FFFFFF}\textbf{\withpatch}} & \multicolumn{9}{c}{\cellcolor[HTML]{FFFFFF}\textbf{\nopatch}}                                                             \\ \hline
\rowcolor[HTML]{EFEFEF} 
Name      & \textbf{\tool}   & $\text{TF-IDF}_{p}$   & $\text{Openai}_{p}$   & \textbf{\tool}            & $\text{Openai}_{a}$ & BM25    & W2V     & CodeBERT & UniXcoder & SFG     & Blizzard & VulF    \\ \hline
\rowcolor[HTML]{FFFFFF} 
MRR       & \textbf{0.78}                   & 0.59                  & 0.68                  & \textbf{0.58}                            & 0.15                & 0.19    & 0.07    & 0.14     & 0.18      & 0.27    & 0.04     & 0.31    \\ \hline
\rowcolor[HTML]{EFEFEF} 
Recall@1  & \textbf{67.04\%}                & 45.50\%               & 56.16\%               & \cellcolor[HTML]{EFEFEF}\textbf{42.00\%} & 10.32\%             & 10.00\% & 2.00\%  & 3.00\%   & 5.00\%    & 13.83\% & 1.00\%   & 16.50\% \\
\rowcolor[HTML]{FFFFFF} 
Recall@3  & \textbf{86.30\%}                & 66.00\%               & 81.72\%               & \textbf{69.67\%}                         & 16.67\%             & 22.33\% & 6.33\%  & 16.50\%  & 23.50\%   & 39.60\% & 3.40\%   & 48.65\% \\
\rowcolor[HTML]{EFEFEF} 
Recall@5  & \textbf{92.59\%}                & 70.50\%               & 84.70\%               & \cellcolor[HTML]{EFEFEF}\textbf{80.50\%} & 23.81\%             & 27.67\% & 10.67\% & 25.00\%  & 31.00\%   & 51.20\% & 6.70\%   & 64.33\% \\
\rowcolor[HTML]{FFFFFF} 
Recall@10 & \textbf{96.30\%}                & 77.00\%               & 84.70\%               & \textbf{84.00\%}                         & 23.81\%             & 35.00\% & 18.17\% & 33.50\%  & 48.50\%   & 62.35\% & 12.00\%  & 75.20\% \\ \hline
\rowcolor[HTML]{EFEFEF} 
ME@1      & \textbf{1.00}                   & 1.00                  & 1.00                  & \cellcolor[HTML]{EFEFEF}\textbf{1.00}    & 1.00                & 1.00    & 1.00    & 1.00     & 1.00      & 1.00    & 1.00     & 1.00    \\
\rowcolor[HTML]{FFFFFF} 
ME@3      & \textbf{1.53}                   & 1.96                  & 1.68                  & \textbf{2.10}                            & 2.77                & 2.15    & 2.93    & 2.87     & 2.23      & 2.12    & 2.99     & 2.06    \\
\rowcolor[HTML]{EFEFEF} 
ME@5      & \textbf{1.76}                   & 2.62                  & 2.02                  & \cellcolor[HTML]{EFEFEF}\textbf{2.88}    & 4.38                & 4.25    & 4.78    & 4.49     & 4.37      & 4.11    & 4.98     & 3.89    \\
\rowcolor[HTML]{FFFFFF} 
ME@10     & \textbf{2.03}                   & 3.96                  & 2.78                  & \textbf{3.47}                            & 8.19                & 7.75    & 9.11    & 8.07     & 7.83      & 6.46    & 10.30    & 6.11    \\
\rowcolor[HTML]{EFEFEF} 
ME@50     & \textbf{3.31}                   & 12.98                 & 8.90                  & \cellcolor[HTML]{EFEFEF}\textbf{8.56}    & 38.67               & 27.98   & 37.06   & 21.58    & 25.42     & 17.55   & 43.75    & 15.44   \\
\rowcolor[HTML]{FFFFFF} 
ME@100    & \textbf{4.88}                   & 24.23                 & 16.55                 & \textbf{12.13}                           & 76.76               & 44.69   & 64.49   & 27.46    & 30.87     & 22.09   & 84.86    & 20.05   \\ \hline
\end{tabular}
\end{table*}
\subsection{RQ2: Ablation Study}

\begin{table}[]
\setlength{\tabcolsep}{1pt}
\footnotesize
\caption{Contribution of Individual Components within \tool in terms of Recall@K, ME@K, and MRR}
\label{tab:ablation}
\begin{tabular}{l|rr|rrrr}
\hline
Mode      & \multicolumn{2}{c|}{\textbf{\withpatch}}                         & \multicolumn{4}{c}{\textbf{\nopatch}}                                         \\ \hline
\rowcolor[HTML]{EFEFEF} 
Name      & \textbf{\tool}            & \textit{w/o \subtool} & \textbf{\tool} & \textit{w/o qe} & \textit{w/o selection} & \textit{w/o rank} \\ \hline
MRR       & \cellcolor[HTML]{FFFFFF}\textbf{0.78}    & 0.72                                 & \textbf{0.58}                 & 0.48            & 0.23                   & 0.10              \\ \hline
\rowcolor[HTML]{EFEFEF} 
Recall@1  & \textbf{67.04\%}                         & 61.10\%                              & \textbf{42.00\%}              & 37.67\%         & 11.83\%                & 6.00\%            \\
Recall@3  & \cellcolor[HTML]{FFFFFF}\textbf{86.30\%} & 78.90\%                              & \textbf{69.67\%}              & 59.83\%         & 24.50\%                & 11.50\%           \\
\rowcolor[HTML]{EFEFEF} 
Recall@5  & \textbf{92.59\%}                         & 83.13\%                              & \textbf{80.50\%}              & 73.00\%         & 36.67\%                & 14.67\%           \\
Recall@10 & \cellcolor[HTML]{FFFFFF}\textbf{96.30\%} & 88.99\%                              & \textbf{84.00\%}              & 79.90\%         & 50.67\%                & 20.33\%           \\ \hline
\rowcolor[HTML]{EFEFEF} 
ME@1      & \textbf{1.00}                            & 1.00                                 & \textbf{1.00}                 & 1.00            & 1.00                   & 1.00              \\
ME@3      & \cellcolor[HTML]{FFFFFF}\textbf{1.53}    & 1.69                                 & \textbf{2.10}                 & 2.21            & 2.70                   & 2.86              \\
\rowcolor[HTML]{EFEFEF} 
ME@5      & \textbf{1.76}                            & 1.96                                 & \textbf{2.88}                 & 2.93            & 4.13                   & 4.61              \\
ME@10     & \cellcolor[HTML]{FFFFFF}\textbf{2.03}    & 3.16                                 & \textbf{3.47}                 & 4.10            & 6.96                   & 8.77              \\
\rowcolor[HTML]{EFEFEF} 
ME@50     & \textbf{3.31}                            & 6.44                                 & \textbf{8.56}                 & 9.30            & 20.41                  & 38.67             \\
ME@100    & \cellcolor[HTML]{FFFFFF}\textbf{4.88}    & 11.38                                & \textbf{12.13}                & 13.504          & 31.10                  & 74.05             \\ \hline
\end{tabular}
\end{table}

To evaluate the three-phrase design of \tool, we created variants by removing individual phases: \textit{w/o \subtool}, \textit{w/o qe}, \textit{w/o selection}, \textit{w/o rank}. In \withpatch mode, \textit{w/o \subtool} is to directly utilize extracted modified methods from patches, highlighting the contribution of \subtool. In the \nopatch mode, each step is iteratively masked out. For example,  \textit{w/o qe} skips QE and keeps other processes identical to \tool. The results are shown in Table~\ref{tab:ablation}. 

As indicated in the left segment of Table~\ref{tab:ablation}, \tool significantly outperforms \textit{w/o \subtool} across all metrics. It achieves an MRR of 0.78 compared to 0.72 by \textit{w/o \subtool}. This difference is primarily because \tool tracks those candidates outside patches to extend the candidate method list through \subtool, whereas \textit{w/o \subtool} is limited to ranking methods directly from patches. The lower MRR and $Recall@K$ for \textit{w/o \subtool} underscore the effectiveness of \subtool in incorporating a broader array of relevant candidate methods. 
Additionally, while $Recall@K$ for \textit{w/o \subtool} approaches 89\% as K increases, it eventually reaches a plateau because it does not cover all relevant methods, being restricted to those within patches. Furthermore, the $ME@K$ for \textit{w/o \subtool} increases more rapidly than for \tool, reflecting a higher likelihood of exhausting the candidate list without successfully identifying the correct \rcfs.

In \nopatch mode, none of the variants match the effectiveness of \tool per analysis:
\begin{itemize}[leftmargin=7pt]
    \item \textbf{\textit{w/o qe}}: Merely using the raw CVE descriptions yields an MRR of 0.48, which is notably lower than \tool's 0.58. This means that QE boosts the effectiveness of \tool by providing more conceptual details. However, in terms of the other two metrics, $Recall@K$ and $ME@K$, \textit{w/o qe} achieves results fairly close to \tool, suggesting that QE primarily enhances the ranking of the first relevant method, with less impact on the overall ranking of all methods.

    \item \textbf{\textit{w/o selection}}: The variant implemented without  \textit{Candidate selection} ranks all non-testing methods directly in the \textit{Ranking} phase. Consequently, the MRR of \textit{w/o selection} is considerably lower than that of \tool, indicating that excessive noise hampers the differentiation of relevant methods during \textit{Ranking}. The other metrics for \textit{w/o selection} also show marked deterioration compared to those of \tool. This underscores the critical role of this step in filtering out irrelevant methods and retaining only potential candidates, thereby enhancing both effectiveness and performance. Moreover, an average of 14,910.03 methods per CVE is reduced to 99.21 methods, a reduction factor of approximately 150 times. It demonstrates this step's ability to eliminate irrelevant methods, which successfully leverages knowledge transferred from \withpatch mode.

    \item \textbf{\textit{w/o rank}}: The absence of the \textit{Ranking} phase results in a low MRR of only 0.10 and inferior other two metrics. This reveals the critical role of \textit{Ranking} in aligning the semantics between the CVE descriptions and the source code, indicating the most significant contribution among steps. Without \textit{Ranking}, the first two steps failed to properly rank the methods, affirming that they are designed to provide only relevant candidates, not to finalize their ranking. 
    Thus, relying solely on the probabilities from \textit{Candidate selection} is not proper. Additionally, we evaluated other commonly used models for Ranking purposes. CodeBert and Word2Vec respectively achieved 0.14 and 0.07 mRR as well as 33\% and 18\% Recall@10. Due to the page limit, the results are displayed on our website.

\end{itemize}
Overall, when all three steps work synergistically, \tool could achieve the best performance.

\begin{tcolorbox}[boxrule=0.5pt,arc=1pt,boxsep=-1mm,breakable]
\textbf{Answer RQ2}: The main components of our \tool framework each play a vital role in enhancing the performance of \rcf localization. In \withpatch mode, \subtool precisely broadens the pool of candidate methods for \textit{Ranking}. In \nopatch mode, three phases prove essential by providing conceptual details, filtering out irrelevant methods, and effectively modeling the correlations, respectively.
\end{tcolorbox}

\subsection{RQ3: Applicability in Real-world Scenarios}
In this RQ, we explored the applicability of \tool in two real-world scenarios regarding the potential boost for SCA tools and new \rcf localization. The evaluations of these two scenarios rely on one dataset introduced afterwards.

\subsubsection{Evaluation Setup}
\ly{To investigate the applicability in real-world cases of \tool, we ran \tool against  
the latest published CVEs in 2024 with different libraries from previously collected ones in the benchmark dataset. As a comparison, $\text{Openai}_{p}$ and VulF as the best baselines for both modes in Section~\ref{sec:rq1} were included.
As of Dec 2024, we collected the latest CVEs designated to Java repositories and fetched as many patch links as possible. As our dataset already covers many commonly used vulnerable Java libraries, there are a limited number of CVEs in 2024 with distinct libraries. In total, 50 CVEs involving 31 distinct libraries were obtained. 
Out of 50 CVEs, 28 were associated with patch links. Separately, we ran \tool against them in \nopatch and \withpatch modes based on the availability of patch links. For 22 CVEs in \nopatch mode, we manually scrutinized the top 10 ranked methods for \rcf. }

The manual verification procedures for CVEs with patches are the same as the ground truth labeling in Section~\ref{sec:label-data}, while the procedures for CVEs without patches are different. 
To ensure reliable labeling, the first two authors separately went through each method and labeled them as \textit{confident} or \textit{likely \rcf}. The methods labeled as \textit{confident} by at least one author were directly marked as \rcf. For \textit{likely \rcf} methods by both, the third author stepped in and made the final decision.
\ly{
However, based on the experience of the three authors, determining if the ranked methods are \rcf is still a subjective task. Thus, besides the majority voting mechanism, we also searched for the potential patches of these 22 vulnerabilities as another source of proof via the Internet or the commit messages in the repository. In total, we found 18 patches for them. Based on the patches, we confirmed the correctness of the \rcf, which served as the ground truths for the evaluation.
}

\begin{figure*}[t!]
    \centering
     \includegraphics[width=0.8\linewidth]{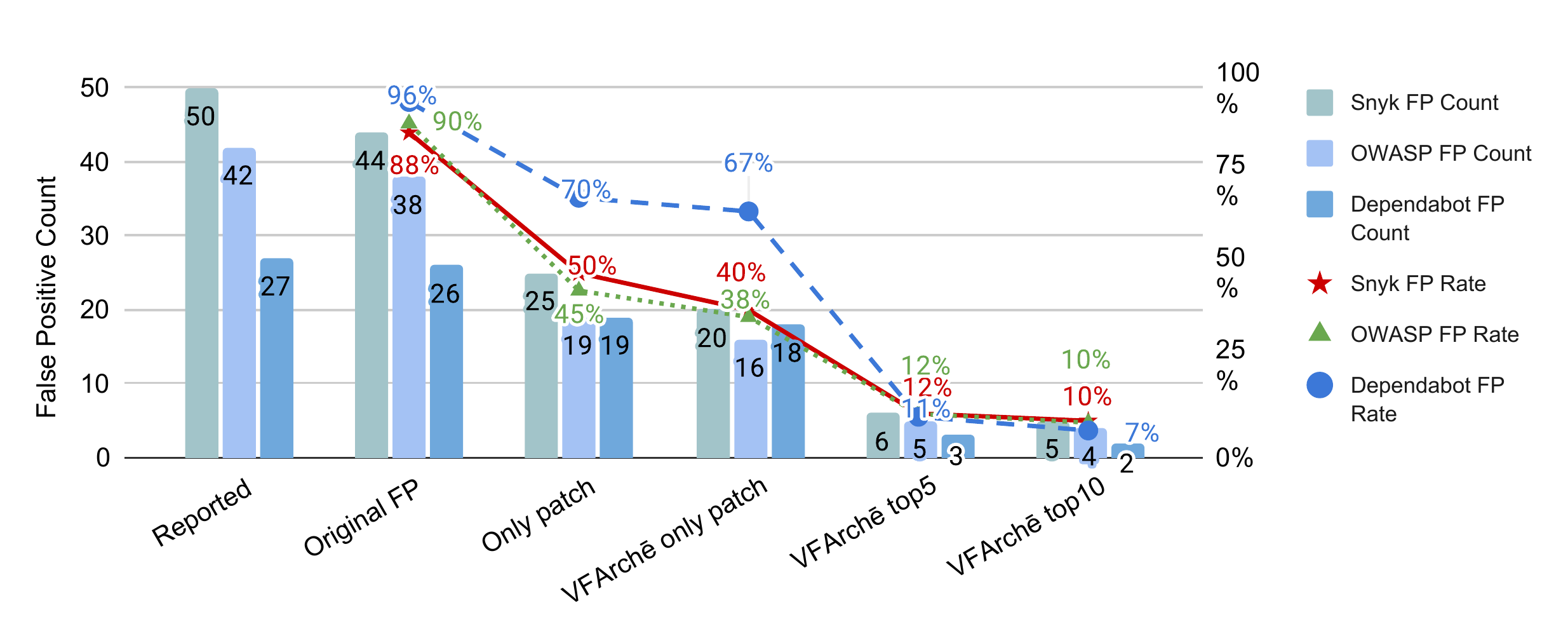}
    \caption{False Positive Reduction for SCA Tools}
    \label{fig:sca}   
\end{figure*}
\subsubsection{Boost for SCA Tools}

\ly{
\rcf serving as critical anchors for the reachability analysis in SCA tools can effectively reduce the false alarms of SCA tools. We experimented to evaluate the boosting effectiveness of the provided \rcf for SCA tools in vulnerability detection based on the latest 2024 CVE dataset in the previous section. From the 31 distinct libraries of 50 CVEs ($L_{vul}$), we could derive another 25 distinct downstream dependents ($L_{user}$) that use the vulnerable libraries from the Maven Central Repository~\cite{mvnrepo}. Then we included both academic and commercial popular SCA tools to detect vulnerabilities within the downstream dependent library set $L_{user}$. The SCA tools include Steady~\cite{steady}, Snyk~\cite{snyk}, OWASP Dependency check~\cite{owaspdependencycheck}, and Dependabot~\cite{dependabot}. However, Steady has its vulnerability database~\cite{steadyvuldata} not updated since 2019, thus being excluded from the evaluation. 
}

\ly{This evaluation aimed to measure how many reported vulnerabilities by SCA tools could be identified as False Positives (FP) by conducting reachability analysis with \rcf.  
The criterion of FP is that the \rcf of a single vulnerability could NOT be reachable from any functions defined in the $L_{user}$ (we conservatively assumed all functions in $L_{user}$ were entry points). The reachability was conducted by Soot~\cite{sootspark} with the SOTA Spark algorithm~\cite{spark}. If \rcf of a vulnerability can be determined by any tools and found unreachable, the FP is eliminated, and then the total FP count declines. 
}

\ly{In Figure~\ref{fig:sca}, the bars indicate the remaining number of FP CVEs before and after applying tools, and the FP rate was calculated by FP counts over the number of reported CVEs. \textit{Reported} is the number of initially reported vulnerabilities by SCA tools. \textit{Original FP} refers to the FP counts prior to applying \tool. \textit{Only patch} refers to the remaining FP counts if only modified functions in available patches were used. \textit{\tool Only patch} refers to using \tool to derive \rcf based on patches while ignoring the \nopatch mode. \textit{\tool top5/10} means using dual modes of \tool to collect \rcf for the reachability, the counts of remaining FP after the elimination with top 5/10 returned functions. As in this dataset, the reachable vulnerabilities were 6, \textit{Original FP} for SCA tools were smaller than \textit{Reported}. Then, with more FPs being determined by providing \rcf, the total FP count plummeted. The remaining FPs were caused by unfound \rcf, resulting in unknown reachability.
}

\begin{tcolorbox}[boxrule=0.5pt,arc=1pt,boxsep=-1mm,breakable]
\textbf{Answer RQ3-1}: Based on 50 CVEs, SCA tools initially reported 88-96\% of FP CVEs. The FP rates of all three tools could be significantly reduced to 10-12\% by applying \tool to localize \rcf for finer-grained SCA vulnerability detection.
\end{tcolorbox}

\subsubsection{New \rcf Localization}
For 28 vulnerabilities with patches, \tool has found \rcf for 27 in the top 10 methods, 24 in the top 5, and 15 in the top 1. Thus results from 27 out of 28 cases could be confirmed by further manual checking. 
Furthermore, we have also labeled whether the patched methods in patches include \rcf. 7 out of 28 CVEs have \rcf in different methods from the patched methods. These 7 vulnerabilities consist of 2 types of scenarios, \textit{Additional Arguments} (4 CVEs) and \textit{Changes of Configurations} (3 CVEs). 
\tool failed at one \textit{Changes of Configurations} case.
In contrast, Baseline $\text{Openai}_{p}$ failed these 5 cases and found 23 in the top 10 due to the limited size of candidate functions.

For the other 22 vulnerabilities without patches, \tool successfully localized \rcf for 16 cases within the top 10 (15 in the top 5). As a comparison, only 5 cases in the top 10 could be localized by VulF, leading to \rcf of 17 cases that cannot be identified with reasonable manual efforts by VulF. Given that most of the cases processed by \tool could be localized for \rcf with reasonable manual efforts (top 5 or 10), \tool could boost the \rcf collection for researchers and industrial practitioners.

\begin{tcolorbox}[boxrule=0.5pt,arc=1pt,boxsep=-1mm,breakable]
\textbf{Answer RQ3-2}: Based on 50 vulnerabilities in 2024 with different vulnerable libraries from previously collected CVEs, \tool successfully localized \rcfs for 43 out of them in the top 10 ranked methods, which is much higher than 26 from baselines. In the real-world scenario, without \tool, relying on baselines, 11 out of 22 \rcfs would fail to be localized in the absence of patches.
\end{tcolorbox}

\section{Discussion Threats to Validity}
The primary threat is the ground truth labeling of the benchmark dataset. Although we have carefully scrutinized the relevant files, it is inevitable to be subjective regarding the \rcf locations. To mitigate this threat, a majority vote was involved.

Another threat is the implementation of baseline tools. Due to the unavailability of artifacts of state-of-the-art tools, such as Ciborowska et al.~\cite{ciborowska2022fast}, we could only use CodeBERT-base and Blizzard for comparison. Our configurations of embedding models follow the general recommendation, but they may not be as optimal as SOTA tools. Moreover, Blizzard performs better in Bug Localization where stack trace data is used. Unfortunately, stack traces are unavailable in CVE descriptions, leading to the sub-optimal performance of Blizzard for the scenario of \rcf localization. More comparisons can be conducted as soon as the SOTA tools release their artifacts.

The last threat is the vulnerable repository mapping with CVEs. NVD and other major vulnerability databases only provide the names of vulnerable libraries. The vulnerable repository mappings with CVEs are not always correct, especially for those libraries comprising multiple repositories. The mapping could be wrong leading to unfound \rcf, which can only be mitigated with a complete repository database.

\section{Related Work}

\subsection{Identifying Missing Information of CVEs}
To augment and enrich the vulnerability-related data for disclosed vulnerabilities like CVEs, plenty of research works have been proposed to identify affected libraries, versions, patches, and so on. For libraries, Chen et al.~\cite{chen2020automated} use the extreme multi-label learning algorithm to identify libraries. Haryono et al.~\cite{haryono2022automated} and Lyu et al.~\cite{lyu2023chronos} have contributed to Chen's approach to accelerate the process. V0Finder~\cite{woo2021v0finder} aims to find the original software with vulnerabilities metadata. For versions, V-SZZ~\cite{bao2022v} is a brand-new SZZ algorithm that leverages the line mapping algorithms to identify the earliest commit. Nguyen et al.~\cite{nguyen2016automatic} scan the
code base and identify the lines of code that were changed to fix the vulnerability. For patches, PatchScout~\cite{tan2021locating} ranks the code commits in the OSS code repository based on their correlations to a given vulnerability.
\ly{For vulnerable files, VulF~\cite{sun2024tracing} proposed a fine-tuned transformer model to trace the vulnerability-relevant files. Besides its granularity of the file level, the model is not suitable for \rcf tracing as proved in our evaluation. 
VFFinder~\cite{wu2024effective} localizes vulnerable functions solely based on CVE descriptions, which neglects the usefulness of patches, thus failing to track \rcf extended from patches like \tool, leading to missing \rcf.
Overall, none of these works effectively in a practical scenario where with or without patches \rcf can always be localized, underscoring the significance of \tool.}

\subsection{Bug Localization}
Bug Localization which aims to localize relevant buggy files/methods commits a similar objective as \rcf localization. The methodologies can be divided into three categories based on the input, namely, the bug report, the stack trace, and the changeset. There are studies focusing on bug report, such as BugScout~\cite{nguyen2011topic}, BLUiR~\cite{saha2013improving}, Ye et al.~\cite{ye2014learning}, BRTracer~\cite{wong2014boosting}, HyLoc~\cite{lam2015combining}, D\&C~\cite{koyuncu2019d}, and BLESER~\cite{zou2021bleser}. Some rely on stack trace, BugLocator~\cite{zhou2012should}, Wang et al.~\cite{wang2013improving}, CrashLocator~\cite{wu2014crashlocator}. A few hybrid tools use both stack traces and bug reports, such as Scaffle~\cite{pradel2020scaffle}, Lobster~\cite{moreno2014use}, and Blizzard~\cite{rahman2018improving}. BLESER~\cite{zou2021bleser} leverages the Abstract Syntax Tree of source code as the semantic representation to localize bugs. There are tools relying on changesets, such as Locus~\cite{wen2016locus}, Ciborowska et al.~\cite{ciborowska2022fast}, and HMCBL~\cite{du2023pre}. Locus~\cite{wen2016locus} was the first tool leveraging changesets to localize bugs. Ciborowska et al.~\cite{ciborowska2022fast} investigates the possibility of tweaking BERT to swiftly localize bugs based on changesets, unlike traditional Bug Localization. HMCBL~\cite{du2023pre} proposes to address the deep semantic representation challenge and the lexical gap between changesets and bug reports. 
Despite numerous works that have studied Bug Localization, they are not suitable for \rcf localization as no structured bug reports or stack trace is available but free-text vulnerability descriptions.

\section{Conclusion}
To advance the security of OSS, we introduced \tool, a framework designed to identify \rcf for published vulnerabilities both when patches are available and when they are not. 
In scenarios where patches are present, we developed \subtool, which utilizes program analysis to mine potential candidate methods. For the more challenging case without patches (\nopatch), we employed knowledge transfer from the \withpatch mode by supervising the fine-tuning of CodeBERT, allowing us to determine a relevant range of candidates.
Candidates are then ranked based on their semantic similarity to expanded descriptions. Our evaluation shows that \tool achieved an MRR of 0.78 in the \withpatch mode and 0.58 in the \nopatch mode. Additionally, an application study confirms that \tool is effective in localizing \rcf for previously unseen CVEs or libraries.
We hope our results can inspire further research in this direction.

\section*{Data Availability}
Our data and source code are available at our website~\cite{dataset}.

\bibliographystyle{IEEEtran}
\bibliography{acmart}

\vfill

\end{document}